\documentclass{INTERSPEECH2023}

 \interspeechcameraready

\usepackage{multirow, tabularx}
\usepackage{pifont}
\usepackage{amssymb}
\usepackage{booktabs}
\usepackage{xcolor}
\usepackage{amssymb}
\usepackage{enumitem}
\setitemize{noitemsep,topsep=0pt,parsep=0pt,partopsep=0pt}
\usepackage{algorithm}
\usepackage{algpseudocode}

\title{Hybrid AHS: A Hybrid of Kalman Filter and Deep Learning for Acoustic Howling Suppression}
\name{Hao Zhang$^1$, Meng Yu$^1$, Yuzhong Wu$^2$, Tao Yu$^2$, Dong Yu$^1$}
\address{
$^1$Tencent AI Lab, Bellevue, WA, USA \\
$^1$Tencent Ethereal Audio Lab, Shenzhen, Guangdong, China}
\email{ \{aaronhzhang@global., raymondmyu@global., yvzhongwu@, taoyuty@, dyu@global.\}tencent.com}

\begin{document}

\maketitle
 
\begin{abstract}
Deep learning has been recently introduced for efficient acoustic howling suppression (AHS). However, the recurrent nature of howling creates a mismatch between offline training and streaming inference, limiting the quality of enhanced speech. To address this limitation, we propose a hybrid method that combines a Kalman filter with a self-attentive recurrent neural network (SARNN) to leverage their respective advantages for robust AHS.
During offline training, a pre-processed signal obtained from the Kalman filter and an ideal microphone signal generated via teacher-forced training strategy are used to train the deep neural network (DNN). During streaming inference, the DNN's parameters are fixed while its output serves as a reference signal for updating the Kalman filter. 
Evaluation in both offline and streaming inference scenarios using simulated and real-recorded data shows that the proposed method efficiently suppresses howling and consistently outperforms baselines.
%


\end{abstract}

\section{Introduction}
%

Acoustic howling is a phenomenon that arises in sound reinforcement systems where the sound emitted from speakers is picked up by a microphone and re-amplified recursively in a feedback loop, resulting in an unpleasant high-pitched sound \cite{waterhouse1965theory, van2010fifty}. This can occur in different settings such as concerts, presentations, public address systems, and hearing aids.
Acoustic howling suppression (AHS) refers to the process of reducing or eliminating the occurrence of acoustic howling. Several methods have been proposed, including passive methods like physical isolation of microphones and speakers, and active methods such as gain reduction \cite{schroeder1964improvement, berdahl2010frequency}, notch filters \cite{loetwassana2007adaptive, gil2009regularized, waterschoot2010comparative}, and adaptive filtering \cite{joson1993adaptive}. 
Among these methods, adaptive filtering, such as the Kalman filter \cite{enzner2006frequency, albu2018hybrid}, dynamically adjusts the signal in real-time to prevent the feedback loop and leads to relatively better speech quality. However, the Kalman filter can be sensitive to control parameters and interferences and fails to address nonlinear distortions introduced by amplifiers and loudspeakers.


Recently, deep learning has been utilized to tackle AHS-related tasks due to its ability to model complex nonlinear relationships. Chen et al. \cite{chen2022neural} introduced a deep learning based method for howling detection. A deep learning based howling noise cancellation method was introduced in \cite{gan2022howling}. Zheng et al. \cite{zheng2022deep} employed deep learning to address the marginal stability problems of acoustic feedback systems, and this method was named as DeepMFC. More recently, a purely deep learning based method (Deep AHS) was proposed for acoustic howling suppression \cite{zhang2023deep}. However, deep learning based methods are prone to a mismatch between offline training and streaming inference, leading to reduced speech quality performance.

Despite significant progress in the development of AHS methods, current methods still face many challenges, especially the trade-off between suppression performance and signal distortion. 
Inspired by the success of combining traditional adaptive methods with deep learning to solve acoustic echo cancellation problems \cite{zhang2022multi, zhang2022deep, zhang2023kalmannet, zhang2022neural}, we present a hybrid method for AHS in this paper.


The proposed method, called Hybrid AHS, combines two approaches to address acoustic howling: a traditional method called frequency domain Kalman filter (FDKF) and a deep neural network (DNN) module based on self-attentive recurrent neural network (SARNN) \cite{yu2022neuralecho}. 
Specifically, the FDKF and SARNN are combined in a cascade manner with the pre-processed output from FDKF serving as an additional input for training the SARNN module. The pre-trained SARNN is then used during streaming inference, and its output is used as a reference signal for updating the FDKF parameters.
During offline training, the Hybrid AHS model is trained in a teacher-forced manner \cite{williams1989learning, lamb2016professor} that assumes only the target speech in the microphone signal is sent to the loudspeaker. This helps convert a recursive howling suppression process to a speech separation problem and shows effective performance for howling suppression during streaming inference.
The proposed method leverages the advantages of both traditional adaptive filtering and deep learning based methods. The benefits of Hybrid AHS are twofold: 1) using the signal pre-processed by traditional method provides more information for model training and helps reduce the mismatch between offline training and streaming inference, and 2) integrating deep learning to further enhance the output of traditional methods resolves the leakages produced due to nonlinear distortion, resulting in a robust solution.


The remainder of this paper is organized as follows. Section 2 introduces acoustic howling problem. The proposed Hybrid AHS method is introduced in Section 3. Section 4 and Section 5 describes the experimental setup and results, respectively. Section 6 concludes the paper.

\section{Acoustic howling suppression}

\begin{figure}[!t]
\centering
     \includegraphics[width=0.99\columnwidth]{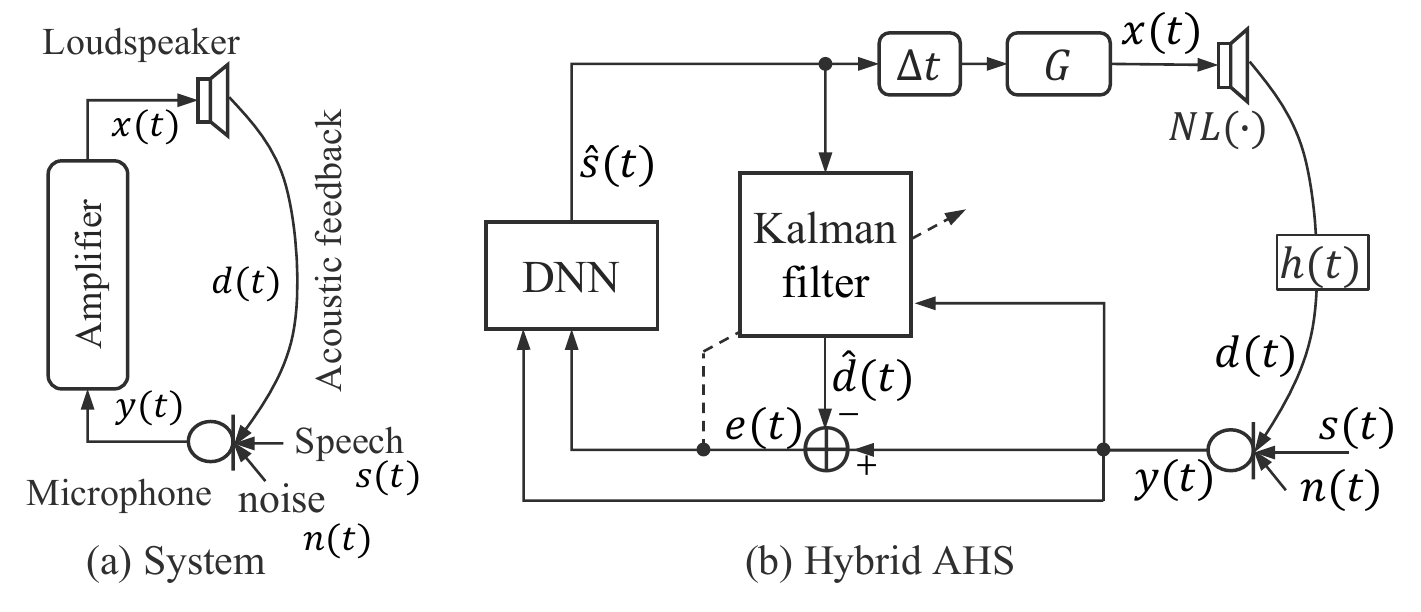}
      \caption{Diagram of (a) an acoustic amplification system, and (b) the proposed Hybrid AHS method.}
      \label{fig:HybridAHS}
\end{figure}

\subsection{Acoustic howling}

A typical single-channel acoustic amplification system is shown in Figure~\ref{fig:HybridAHS}(a). It consists of a microphone and a loudspeaker where the target speech is picked up by the microphone as $s(t)$, which is then sent to the loudspeaker for acoustic amplification. The loudspeaker signal $x(t)$ is played out and arrives at the microphone as an acoustic feedback denoted as $d(t)$:
\begin{flalign}
d(t) = \textstyle NL[x(t)]*h(t)
\end{flalign}
where $NL(\cdot)$ denotes the nonlinear distortion introduced by the loudspeaker, $h(t)$ represents the acoustic path from loudspeaker to microphone, and $*$ denotes linear convolution. 

If without any processing, the playback signal $d(t)$ will re-enter the pickup repeatedly, the corresponding microphone signal can then be represented as:
\begin{flalign}
\label{equ:howling}
\textstyle y(t) =   s(t) + n(t) +  NL\left[y(t-\Delta t) \cdot G\right]*h(t) 
\end{flalign}
where $n(t)$ represents the background noise, $\Delta t$ denotes the system delay from microphone to loudspeaker, $G$ is the gain of amplifier.
The recursive relationship between $y(t)$ and $y(t-\Delta t)$ causes re-amplifying of playback signal and leads to a feedback loop that results in an annoying, high-pitched sound, which is known as acoustic howling. 

It is worth acknowledging that acoustic howling and acoustic echo are two distinct phenomena, although inappropriate handling of acoustic echo can result in howling. The primary differences between these two phenomena are: 1. While both of them are fundamentally playback signals, howling is characterized by a gradual buildup of signal energy in a recursive manner. 2. The signal that leads to howling is generated by the same source as the target signal, making the suppression of howling more challenging.

\subsection{Existing AHS methods}

In general, the goal of an AHS is to reduce or eliminate the howling, while preserving the desired signal as much as possible. 
However, this goal is not always easy to achieve, because the suppression of the howling can often result in some level of signal distortion.  
Kalman filter based methods have a long history of success in a variety of signal processing applications, and they can be effective in suppressing acoustic howling in certain environments \cite{albu2018hybrid}. 
However, they are limited by their reliance on a statistical model of the system, which can be difficult to estimate accurately in scenarios with nonlinearities and leads to noticeable leakage and/or signal distortion.
Deep learning based methods, on the other hand, can learn complex relationships and be effective in suppressing acoustic howling in environments where the howling can not be well modeled by a Kalman filter \cite{zhang2023deep}. While the mismatch between offline training and streaming inference leads to unavoidable signal distortions.

The challenge is to find a balance between suppression performance and signal distortion that is acceptable for a particular application.
Therefore, we consider combine traditional and deep learning based methods to achieve improved performance, improved robustness, and the ability to leverage the strengths of both methods to achieve a more effective solution.

\section{Proposed method: Hybrid AHS}

\begin{figure*}[!t]
\centering
     \includegraphics[width=1.8\columnwidth]{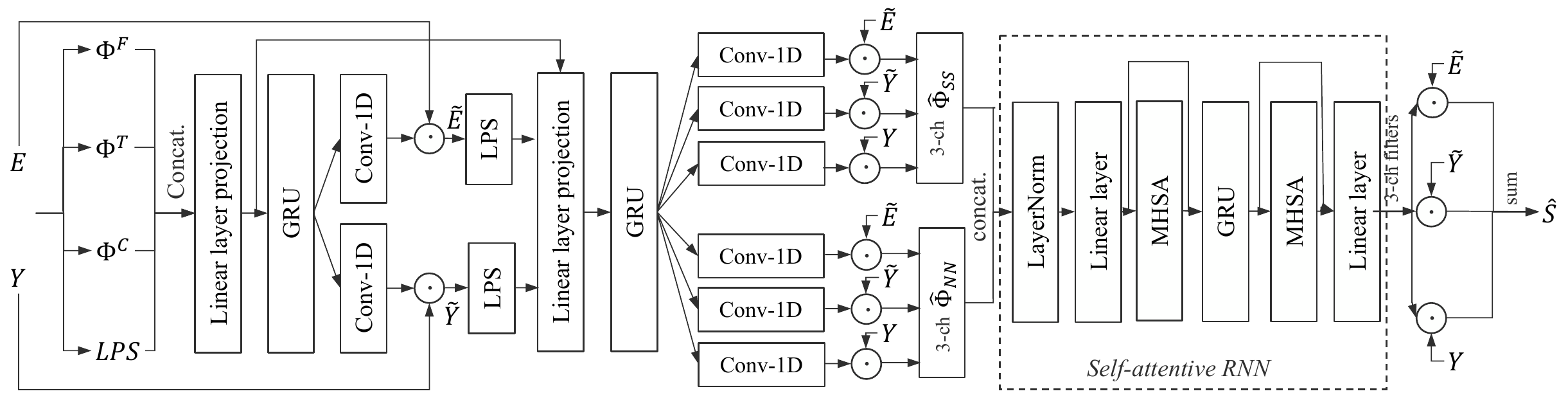}
      \caption{Architecture of the DNN module for AHS. Each ``Conv-1D'' block outputs a complex-valued ratio filter, which is then applied upon a specific signal through deep filtering, denoted as $\odot$. }
      \label{fig:NeuralEcho}
\end{figure*}

To address the disadvantages of the above mentioned AHS methods and leverage their advantages, a hybrid solution, called Hybrid AHS, is proposed in this paper. 
Figure~\ref{fig:HybridAHS} (b) illustrates a schematic of the proposed method, which comprises of two modules: Kalman and DNN.


\subsection{Problem formulation}


Suppressing howling is best achieved by incorporating the AHS method within the acoustic loop considering the recursive nature of howling.
However, this can be computationally demanding and inefficient for deep learning based methods. 
To address this challenge, we follow the approach of Deep AHS \cite{zhang2023deep} and adopt the teacher-forcing training strategy to formulate AHS as a speech separation problem during model training. 
This proposed approach is based on the assumption that the Hybrid AHS model, once properly trained, can attenuate interferences and transmit only the target speech to the loudspeaker. 
Consequently, the actual output $\hat{s}(t)$ in Figure~\ref{fig:HybridAHS}(b) can be replaced with the ideal target (teacher signal) $s(t)$ during model training, and the recursively defined microphone signal in equation (\ref{equ:howling}) is converted into a mixture of target signal, background noise, and an one-time playback signal determined by $s(t)$:
\begin{flalign}
\label{equ:howling2}
\textstyle y(t) =   s(t) + n(t) +  h(t)*NL[s(t-\Delta t)\cdot G]
\end{flalign}
The overall task of AHS is then transformed into a speech separation problem during offline training. 
The object is to extract the target signal $s(t)$ from the ideal microphone signal, defined in equation (\ref{equ:howling2}) and exclusively employed for model training, using the Kalman filter output $e(t)$ as an additional input, thus jointly suppressing howling and noise.



\subsection{Kalman filter}

The Kalman module utilizes microphone signal $y(t)$ and the enhanced signal $\hat{s}(t)$ as a reference (denoted as $r(t)$) to obtain an estimate of the acoustic path $\hat{h}(t)$ and the corresponding feedback $\hat{d}(t)$. The estimated feedback is then subtracted from the microphone signal, and the resulting error signal $e(t)$ is employed for weight updating. 
The overall process can be viewed as a two-step procedure (prediction and updating) with Kalman filter weights updated through the iterative feedback from the two steps. 

In the prediction step, the near-end signal is estimated as 
\begin{equation}
    {\mathbf{E}}(k) = \mathbf{Y}(k) - \mathbf{R}(k)\hat{\mathbf{H}}(k),
\end{equation}
where ${\mathbf{E}}$, $\mathbf{Y}$, and $\mathbf{R}$ are the short-time Fourier transform (STFT) of $e(t)$, $y(t)$, and $r(t)$ respectively, and $k$ denotes the frame index. $\hat{\mathbf{H}}(k)$ denotes the frequency-domain estimated echo path. 

The echo path $\hat{\mathbf{H}}(k)$ is updated in the updating step: 
\begin{equation}
    \label{eq:state}
    \hat{\mathbf{H}}(k+1) = A[\hat{\mathbf{H}}(k)+\mathbf{K}(k){\mathbf{E}}(k)],
\end{equation}
where $A$ is the transition factor. $\mathbf{K}(k)$ denotes the Kalman gain, which is obtained using covariances calculated from state estimation error, observation and process noises \cite{enzner2006frequency}. 

\subsection{Inputs and feature extraction}

The DNN module, illustrated in Figure~\ref{fig:NeuralEcho}, accepts a pre-processed signal using the Kalman fitler $e$ and an ideal microphone signal generated via teacher forcing learning $y$  as inputs for model training.
The input signals, which are sampled at 16 kHz, are split into frames of 32 ms and a frame shift of 16 ms. A 512-point STFT is then performed on each frame, resulting in the frequency domain inputs, $Y$ and $E$.
Besides the normalized log-power spectra (LPS), we extract the correlation matrix across time frames and frequency bins of the input signals to capture the signals' temporal and frequency dependency.  These features help in differentiating between howling and tonal components.
Channel covariance of input signals ($Y$ and $E$) is calculated as another input feature to account for cross-correlation between them. 
A concatenation of these features is used for model training with a linear layer for feature fusion. 
More details regarding feature design can be found in \cite{yu2022neuralecho}


\subsection{Network structure}

The DNN module is implemented using a self-attentive recurrent neural network (SARNN). 
The neural network is composed of three main parts. 
The first part comprises a gated recurrent unit (GRU) layer with 257 hidden units and two 1D convolution layers. These layers estimate two complex-valued filters which are applied on the input signals using deep filtering \cite{mack2019deep} to obtain intermediate outputs, denoted as $\tilde{Y}$ and $\tilde{E}$. The motivation behind obtaining these intermediate outputs is that they can be used as learnt nonlinear reference signals \cite{zhang2022deep, zhang2023kalmannet} and provide more information for howling suppression. 
Later, the LPS of these intermediate signals are concatenated with the fused feature and then used as inputs for another GRU layer. We regard $Y$, $\tilde{Y}$, and $\tilde{E}$ as three-channel inputs and employ two 1D convolution layers for each input channel to estimate the playback/noise and target speech components in it. The corresponding covariance matrices of playback/noise $\hat{\Phi}_{NN}$ and target speech $\hat{\Phi}_{SS}$ are calculated and concatenated as the input to the third part, SARNN. 
The SARNN part employs two linear layers, two multi-head self-attention (MHSA), a GRU, and residual connections to estimate a three-channel enhancement filter. The enhanced signal $\hat{S}$ is then obtained through multi-channel deep filtering. 
Finally, an inverse STFT (iSTFT) is used to get waveform $\hat{s}$. Details of the network structure can be found in \cite{yu2022neuralecho}.


\subsection{Loss functions}
We utilize a combination of scale-invariance signal-to-distortion ratio (SI-SDR) \cite{le2019sdr} in the time domain and mean absolute error (MAE) of spectrum magnitude in the frequency domain for model training:
\begin{flalign}
Loss = -\text{SI-SDR}(\hat{s}, s) + \lambda \text{MAE} (|\hat{S}|, |S|)
\end{flalign}
$\lambda$ is set to 10000 to balance the value range of these two losses.

\begin{figure*}[!t]
\centering
     \includegraphics[width=2.1\columnwidth]{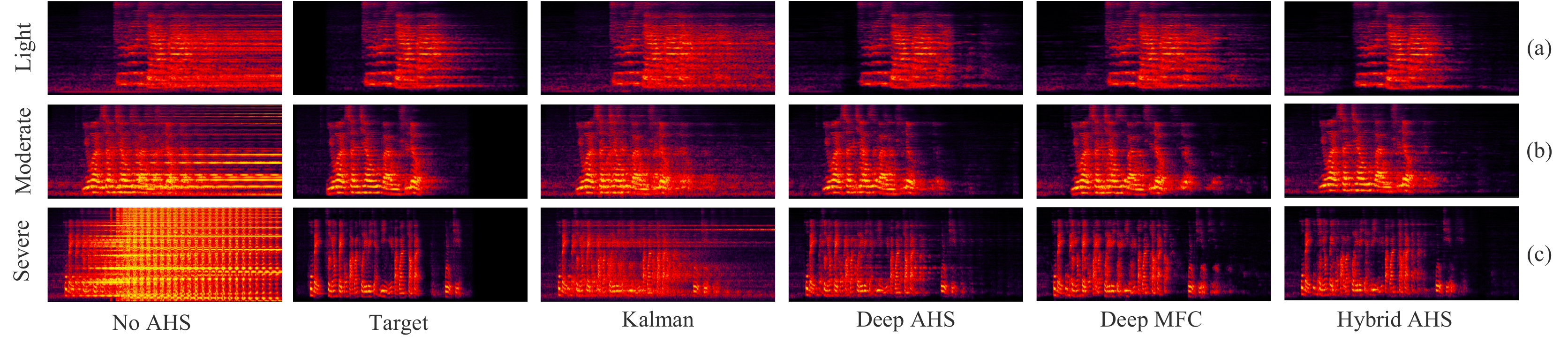}
      \caption{Spectrograms of streaming results for scenarios with (a) light, (b) moderate, and (c) severe howling. The signals with no AHS, target, Kalman, Deep AHS, Deep MFC, and the proposed Hybrid AHS are shown in each panel.}
      \label{fig:spectrogram}
\end{figure*}

\section{Experimental setup}

\subsection{Data preparation}

The AISHELL-2 dataset \cite{du2018aishell} is used for carrying out experiments in situations with playback, background noise, and nonlinear distortions. 
During data simulation, we generate 10,000 room impulse response (RIR) sets using the image method \cite{allen1979image} with random room characteristics and reverberation times (RT60) range of 0 to 0.6 seconds. Each RIR set consisting of RIRs for the near-end speaker, loudspeaker and background noise locations. 
A randomly selected RIR set is utilized to generate target speech $s(t)$ and its corresponding one-time playback signal $d(t)$ using system delay $\Delta t$ randomly generated within the range of $[0.1, 0.3]$ seconds, and amplification gain randomly selected within the range of $[1, 3.2]$. The nonlinear distortions introduced by the amplifier and loudspeaker are simulated as a saturation type of nonlinearity using hard clipping and Sigmoidal function \cite{birkett1996nonlinear, zhang2018deep}.
The microphone signal for offline training is created as a mixture with signal-to-playback-ratio (SPR) randomly selected from $[-10, 10]$ dB and signal-to-noise ratio (SNR) ranging from $-10$ dB to $30$ dB. Kalman filter is employed in an initial stage to preprocess the training signals and obtain the corresponding outputs. 
A total number of 10k, 0.3k, and 0.5k signals are generated for training, validation and testing, respectively. The testing data are generated using different utterances and RIRs from that of training and validation data.  The model is trained for 60 epochs with a batch size of 32.

\subsection{Evaluation metrics}

The performance of the proposed method is evaluated in two different manners: offline evaluation and streaming inference \cite{zhang2023deep}.
The offline evaluation uses signals generated in Eq.~(\ref{equ:howling2}) as input to evaluate playback attenuation performance. SI-SDR and perceptual evaluation of speech quality (PESQ) \cite{rix2001perceptual} are used to evaluate the extent of playback attenuation and quality of target speech. And a higher value denotes better performance. 


In streaming inference, we insert the deep learning module into the acoustic loop and generate the enhanced signal recursively. This manner of evaluation considers the potential re-entry of leakage/distortion in the close acoustic loop and evaluates the proposed method's real-time howling suppression performance \cite{zhang2023deep}. The spectrograms of recursively processed signal are presented to show the effectiveness of the proposed method.

\section{Experimental results}

\subsection{Offline evaluation}

\begin{table}[!t]
\centering
\caption{Offline evaluation of models for playback attenuation.}
\label{table:input}
\resizebox{0.48 \textwidth}{!}{
\begin{tabular}{l|ccc|ccc} \specialrule{1.5 pt}{1 pt}{1 pt}  \hline
               & \multicolumn{3}{c|}{SI-SDR (dB)} & \multicolumn{3}{c}{PESQ} \\ \hline
$G$       &    1       &     2     &     3     &   1     &     2  &  3      \\  \hline                 
Unprocessed  &  8.59     &   2.82        &    -0.66         & 2.83   & 2.41       &  2.18       \\
Kalman  & 7.66    &    -1.42       &      -10.56    & 2.88   &   2.44     &    2.07       \\
Deep AHS \cite{zhang2023deep}   & 16.32     &  13.32     &   11.46      & 3.60    &  3.41    &   3.27    \\
Deep MFC \cite{zheng2022deep}  &  14.82  &   8.60      &   2.61     &  3.46     & 3.13      &    2.81       \\
Hybrid AHS & \textbf{20.16} & \textbf{17.11}    & \textbf{14.43}  &   \textbf{3.76}   &  \textbf{3.60}      &   \textbf{3.43}      \\
 \hline
\specialrule{1.5 pt}{1 pt}{1 pt}
\end{tabular}
}
\end{table}

We first evaluate the performance of our proposed method for playback attenuation and compare it with two recently proposed deep learning based AHS methods \cite{zheng2022deep, zhang2023deep} and a Kalman filter based approach \cite{albu2018hybrid}. The comparison results are presented in Table~\ref{table:input}, with $G$ representing the amplification gain. 
Note that the Kalman results shown here could be even worse than the "Unprocessed" signals since the former is obtained in a streaming manner, whereas the latter is an ideal microphone mixture generated using the teacher-forced training strategy described in Eq.~(\ref{equ:howling2}). 
To ensure a fair comparison, we trained the Deep AHS, Deep MFC, and Hybrid AHS using the same network and training data. The comparison results indicate that the Hybrid AHS approach outperforms all baseline methods consistently.
We have also experimented with using a delayed microphone signal as another reference signal, as suggested in \cite{zhang2023deep}. However, our findings indicate that incorporating a delayed microphone does not lead to performance improvement for Hybrid AHS, since the output of Kalman  provides sufficient reference information and outperforms the benefits of having a delayed microphone as additional input.

\subsection{Streaming inference}



This section assesses the effectiveness of the proposed method using streaming inference. Three testing scenarios, comprising soft, moderate, and severe howling, are generated by gradually increasing the amplification gain $G$ during streaming inference. The spectrograms of processed signals are shown in Figure~\ref{fig:spectrogram}.
It can be observed that all deep learning based AHS approaches successfully prevent the occurrence of howling, with the Hybrid AHS method delivering the best overall performance.

\subsection{Performance using real-recorded signals and a deployable model}

\begin{figure}[!t]
\centering
     \includegraphics[width=0.9\columnwidth]{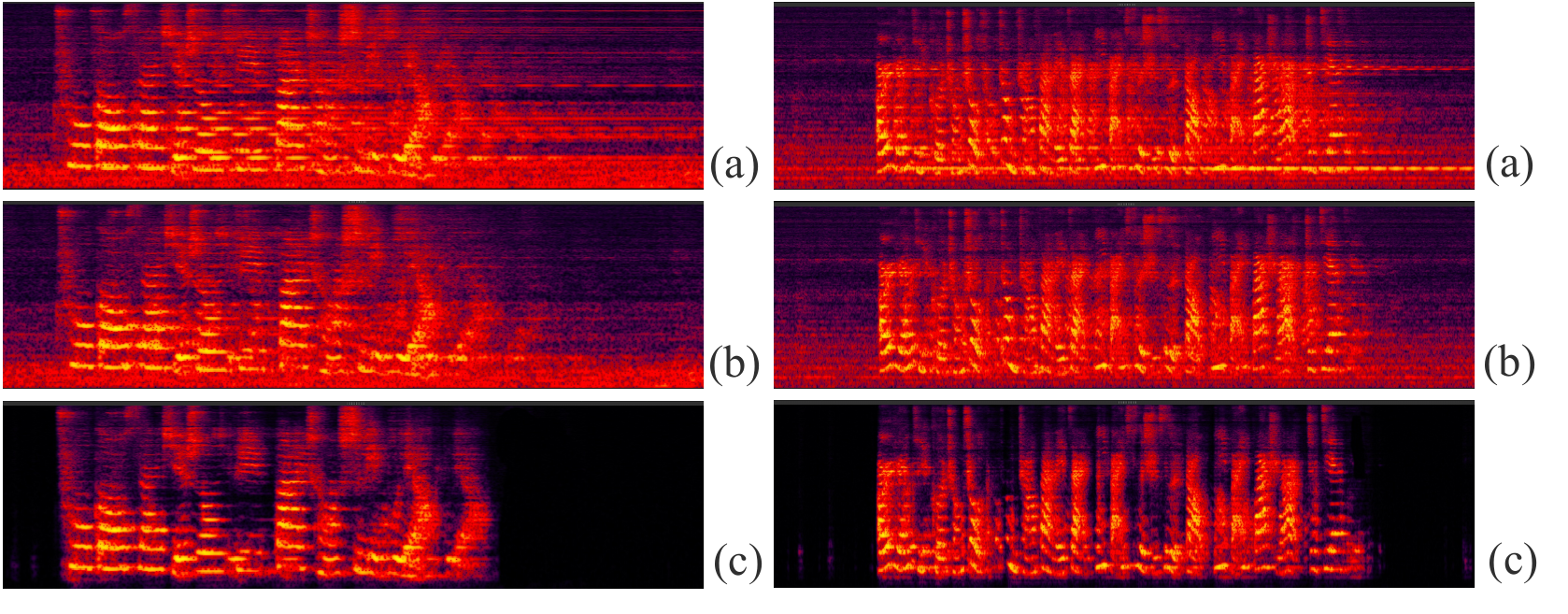}
      \caption{Spectrograms of two test results using real recordings and a small model when (a) turn off both Kalman and DNN modules, (b) turn on only Kalman module, and (c) turn on both Kalman and DNN modules (proposed Hybrid AHS).}
      \label{fig:real}
\end{figure}

We further evaluate the performance of the proposed method for howling suppression with real-recorded signals and a deployable model. 
For this purpose, we set up a simple acoustic amplification system and trained the Hybrid AHS method using real recordings and a small, deployable model. The model is a long short-term memory (LSTM) network that consists of a single hidden layer with 100 units, resulting in 0.13 M trainable parameters. To make the model feasible for deployment on real devices, we reduced the frame size and frame shift to 8 ms and 4 ms, respectively, and used only LPS features as inputs for model training.
The processed results are presented in Figure~\ref{fig:real}, demonstrating the effectiveness of the proposed method in howling suppression and its ability to further enhance the output of the Kalman filter.

\section{Conclusion}

In this study, we have introduced a Hybrid AHS approach that integrates traditional Kalman filtering with deep learning to suppress acoustic howling. The proposed method involves offline training of a SARNN using signals that have been pre-processed by Kalman filtering, as well as a microphone signal generated using teacher forcing training strategy. During streaming inference, the pre-trained model is inserted into the closed acoustic loop to recursively process the input signals. By leveraging both Kalman filtering and deep learning, the proposed method achieves enhanced suppression performance and speech quality for nonlinear AHS in comparison to baseline techniques in both offline and streaming scenarios. 
Future work includes exploring practical issues such as on-device implementation and extending the proposed method for handling multi-channel scenarios.

\bibliographystyle{IEEEtran}
\bibliography{Hybrid_AHS}

\end{document}